\documentstyle[12pt]{article}

\newcommand{\beq}{\begin{equation}}
\newcommand{\eeq}{\end{equation}}
\newcommand{\beqa}{\begin{eqnarray}}
\newcommand{\eeqa}{\end{eqnarray}}
\newcommand{\beqar}{\begin{eqnarray*}}
\newcommand{\eeqar}{\end{eqnarray*}}

\newcommand{\Ga}{\Gamma}
\newcommand{\ka}{\kappa}
\newcommand{\inn}{\!\!\cdot\!\!}

\renewcommand{\l}{\lambda}

\newcommand{\sig}{\sigma}

\newcommand{\z}{\zeta}

\newcommand{\eg}{{\it e.g.,}\ }
\newcommand{\ie}{{\it i.e.,}\ }
\newcommand{\labell}[1]{\label{#1}} 
\newcommand{\reef}[1]{(\ref{#1})}
\newcommand\prt{\partial}
\newcommand\veps{\varepsilon}
\newcommand\ls{\ell_s}
\newcommand\cF{{\cal F}}
\newcommand\cL{{\cal L}}
\newcommand\cG{{\cal G}}

\newcommand\bz{\bar{z}}
\newcommand\bw{\bar{w}}
\newcommand\hF{\hat{F}}
\newcommand\hA{\hat{A}}
\newcommand\hD{\hat{D}}
\newcommand\hf{\hat{f}}
\newcommand\ha{\hat{a}}
\newcommand\hl{\hat{\lambda}}

\newcommand\tG{{\widetilde G}}

\newcommand\tphi{{\tilde \phi}}
\newcommand\tpsi{{\tilde \psi}}
\newcommand\tX{{\tilde X}}
\newcommand\tB{{\widetilde B}}

\newcommand\tV{{\widetilde V}}
\newcommand\Tr{{\rm Tr}}

\begin{document}

\thispagestyle{empty}
\rightline{\small hep-th/9909214 \hfill IPM/P-99/051}
\vspace*{1cm}

\begin{center}
{\bf \Large 
Non-commutative world-volume interactions on D-brane \\[.25em]
and Dirac-Born-Infeld action  }
\vspace*{1cm}

{Mohammad R. Garousi\footnote{E-mail:
biruni@iran.com}}\\
\vspace*{0.2cm}
{\it Department of Physics, University of Birjand, Birjand, Iran}\\
\vspace*{0.1cm}
and\\
{\it Institute for Studies in Theoretical Physics and Mathematics IPM} \\
{P.O. Box 19395-5746, Tehran, Iran}\\
\vspace*{0.4cm}

\vspace{2cm}
ABSTRACT
\end{center}
By integrating the  Seiberg-Witten differential equation in a special path, 
we write ordinary gauge
fields in terms of their non-commutative counterparts up to three 
non-commutative gauge fields. 
We then use this change of variables  to
write ordinary abelian Dirac-Born-Infeld action in terms of 
non-commutative fields.
The resulting action is then compared  with various low energy contact terms
of world-sheet perturbative string scattering amplitudes from 
non-commutative D$p$-brane. We find 
completely agreement between the field theory and string theory results.
Hence, it shows that perturbative string theory knows the solution of
the Seiberg-Witten differential equation.
\vfill
\setcounter{page}{0}
\setcounter{footnote}{0}
\newpage

\section{Introduction} \label{intro}

Recent years have seem dramatic progress in the understanding
of non-perturbative aspects of string theory\cite{excite}.
With these studies has come the realization that extended
objects, other than just strings, play an essential role.
An important tool in these investigations has been Dirichlet
branes\cite{joep}. D-branes are non-perturbative states on which open
string can live, and to which various closed strings including
Ramond-Ramond states can couple.

Another interesting aspect of D-branes is that in the presence of
background flux the world-volume of D-brane becomes non-commutative\cite{noncom1,noncom2,noncom3}.
Hence, at low energy the D-brane dynamics may be described by non-commutative
gauge theory. On the other hand, it is known that the D-brane  is 
properly described  by Dirac-Born-Infeld action with appropriate  background
flux (see \eg \cite{newts}). Using this idea, 
Seiberg and Witten were able to find, among other things,
an explicit
differential equation  that relates non-commutative gauge fields at different
non-commutative parameter\cite{sw}. This Seiberg-Witten differential equation
can be integrated  to find a transformation that changes ordinary
gauge fields into non-commutative fields.

The purpose  of this paper 
is  to show that the world-sheet perturbative string theory can capture above
transformation.  
To this end,
we  
integrate the differential equation  in a special path  to write ordinary
gauge fields in terms  of their
non-commutative counterparts up to three non-commutative fields. The resulting
transformation for abelian case contains two different  multiplication rules. One
is the familiar non-commutative $*$ multiplication that appears 
in the definition of non-commutative
gauge field strength and the other one, we call $*'$ multiplication 
(see \reef{star'}),
 operates
as commutative  multiplication rule between two non-commutative field strengths.
We 
use this transformation to rewrite the ordinary DBI action in terms of non-commutative
fields. The resulting field theory action which
contains various new interactions between ordinary closed string  
and non-commutative open string fields are then compared with approperaite 
low energy
contact terms of perturbative string theory. 
Our results in string theory side
are fully consistent with
the new interactions in field theory side. Hence, 
it shows that perturbative string theory knows about the solution of the
Seiberg-Witten differential equation.

One of the outcome of our calculations is that if one only replaces
ordinary fields in the  DBI action
in terms of their non-commutative counterparts, the resulting action is not completely
identical to  the  approperaite contact terms of string theory scattering amplitudes. 
To have
an action that is fully compatible with string theory, one should transform 
ordinary multiplication rule between open string fields to the $*'$ rule as well. 
  
The paper is organized as follows. In the following section we expand the
DBI action to produce various interactions involving one closed and one or
two ordinary open string fields. In Section 3, 
we integrate the Seiberg-Witten differential equation
to  transform the ordinary open string fields to their non-commutative 
counterparts. This transformation leads us to propose that
the ordinary multiplication rule between two open string fields 
in the expansion of DBI action should be replaced by 
the commutative $*'$ rule. 
In Section 4, using the conformal field theory technique, we calculate various 
string theory amplitudes describing scattering of closed and open string states
from the non-commutative D$p$-branes. Using these string theory amplitudes,
we determine various low energy amplitudes and contact terms and compare them
with the field theory results.
We conclude with a brief discussion of our results in Section 5. 
Appendix contains our conventions and some useful comments  
on conformal field theory propagators and
vertex operators used in our calculations.

\section{Dirac-Born-Infeld Couplings}

The world-volume theory of a single D-brane in type 0 theory 
includes a massless
U(1) vector $A_a$ and  a set of massless scalars $X^i$, describing the transverse
oscillations of the brane
\cite{leigh,kelone}. The leading order low-energy action for these
fields corresponds to a dimensional reduction of a ten dimensional
U(1) Yang Mills theory. As usual in string theory, there are
higher order $\alpha'=\ls^2$ corrections,
where $\ls$ is the string length scale. As long as derivatives
of the field strengths (and second derivatives of the scalars)
are small compared to $\ls$, then the action takes a Dirac-Born-Infeld
form \cite{bin}. 
To take into account the couplings of the open string states
with closed strings, the DBI
 action may be extended naturally to include background
closed string fields, in particular, the metric, dilaton, 
Kalb-Ramond and tachyon field\cite{type0,keltwo}. In this case  one arrives at the 
following world-volume
action:
\beq
S_{BI}=-T_p \int d^{p+1}\sig\ g(T)e^{-\Phi}\sqrt{-det(\tG_{ab}+
\tB_{ab}+2\pi\ls^2\,F_{ab})}
\labell{biact}
\eeq
where the tachyon function is $g(T)=1+T/4+3T^2/32+\cdots$ \cite{type0}.
Here, $F_{ab}$ is the abelian field
strength of the world-volume ordinary gauge field, while
the metric and antisymmetric tensors are
the pull-backs of the bulk tensors to the D-brane world-volume, \eg
\beq
\tG_{ab}=G_{ab}+2G_{i(a}\,\prt_{b)}X^i+G_{ij}\prt_aX^i\prt_bX^j\ .
\labell{pull}
\eeq
In general, the closed string fields are function of world-volume and
transverse coordinates, \ie $X^a$ and $X^i$ respectively, however, for simplicity we assume
they are just function of $X^a$. 

In order to find  the interactions expected from the  DBI 
action, we expand  
the action for fluctuations around 
$G_{\mu\nu}=\eta_{\mu\nu}$, $B_{\mu\nu}=\cF^{ab}\eta_{a\mu}\eta_{b\nu}$,
$\Phi=0$. The fluctuations
should be normalized as the conventional field theory modes which
appear in the string vertex operators. As a first step, we recall
that the graviton vertex operator corresponds to string frame metric. Hence, one
should transform the Einstein frame metric $G_{\mu\nu}$ to the string frame
metric $g_{\mu\nu}$ via
$G_{\mu\nu}=e^{\Phi/2}g_{\mu\nu}$.
Now with conventions of \cite{ours}, the string mode fluctuations take the
form
\beqa
g_{\mu\nu}&=&\eta_{\mu\nu}+2\kappa h_{\mu\nu}\nonumber\\
\Phi&=&\sqrt{2}\kappa\phi\nonumber\\
B_{\mu\nu}&=&\cF^{ab}\eta_{a\mu}\eta_{b\nu}-2\kappa b_{\mu\nu}
\nonumber\\
T&=&2\kappa\tau\nonumber\\
A_a&=&{1\over\sqrt{T_p}2\pi\ls^2}a_a\nonumber\\
X^i&=&{1\over\sqrt{T_p}}\l^i\,\, .
\labell{onormal}
\eeqa
With these normalizations, the pull back of the Einstein frame metric becomes:
\beqa
\tG_{ab}&=&\eta_{ab}(1+\frac{\ka}{\sqrt{2}}\phi)+2\ka{\tilde{h}}_{ab}+
\frac{1}{T_p}
(1+\frac{\ka}{\sqrt{2}}\phi)\prt_a\l^i\prt_b\l_i+\cdots
\eeqa
where the dots represents terms with two and more closed string fields.  

Now  it is straightforward,
to expand eq.~\reef{biact} using
\beqar
\sqrt{det(M_0+M)}\!\!\!&=&\!\!\!\sqrt{det(M_0)}(
1+{1\over2}\Tr(M_0^{-1}M)
-{1\over4}\Tr(M_0^{-1}MM_0^{-1}M)+{1\over8}(\Tr(M_0^{-1}M))^2
\nonumber\\
&&
+{1\over6}\Tr(M_0^{-1}MM_0^{-1}MM_0^{-1}M)-{1\over8}\Tr(M_0^{-1}M)
\Tr(M_0^{-1}MM_0^{-1}M)\nonumber\\
&&+
{1\over48}(\Tr(M_0^{-1}M))^3+\ldots)
\eeqar
to produce a vast array of interactions. We are mostly interested in
the interactions linear in the closed string fluctuations, and
linear or quadratic in the open string fields. 

We begin with the linear couplings of the closed strings to the
D-brane source itself 
\beqa
\cL_{0,1}&=&-T_p\ka c\left(\frac{1}{2}\tau+V^{ab}(h_{ba}-b_{ba})+
\frac{1}{2\sqrt{2}}
(\Tr(V)-4)\phi\right)
\labell{int0}
\eeqa
where we defined the  overall square root of the metric as 
$\sqrt{-det(\eta_{ab}+\cF_{ab})}\equiv c$, and
matrix $V^{ab}$ as the dual of the metric, that is
\beqa
V^{ab}&\equiv &\left( (\eta+\cF)^{-1}\right)^{ab} \,\, .
\labell{vmatrix}
\eeqa
Next there are interactions involving one closed string mode and
one open string mode, that is 
\beqa
\cL_{1,1}&=&-\sqrt{T_p}\ka c\left(\frac{1}{2}V^{ab}f_{ba}
(\frac{1}{2}\tau+V^{ab}(h_{ba}-b_{ba})+\frac{1}{2\sqrt{2}}
(\Tr(V)-4)\phi)\right.\nonumber\\
&&\left.-V^{ab}(h_{bc}-b_{bc}+\frac{1}{2\sqrt{2}}\phi\eta_{bc})V^{cd}f_{da}
+2V^{ab}(h_{i(b}\prt_{a)}\l^i
-b_{i[b}\prt_{a]}\l^i)\right)
\labell{int1}
\eeqa
where $f_{ab}=\prt_aa_b-\prt_ba_a$. 
We will need also to compare our results with the DBI terms
that have one closed and two open string states, 
\beqa
\cL_{2,1}&=&-\ka c\left(({1\over2}\tau+V^{ab}(h_{ba}-b_{ba})+
\frac{1}{2\sqrt{2}}(\Tr(V)-4)\phi
)\right.\nonumber\\
&&\times({1\over2}V^{ab}\prt_a\l^i\prt_b\l_i-{1\over4} V^{ab}
f_{bc}V^{cd}f_{da}
+{1\over8}(V^{ab}f_{ba})^2)\nonumber\\
&&-V^{ab}(h_{bc}-b_{bc}+\frac{1}{2\sqrt{2}}\phi\eta_{bc})
(V^{cd}\prt_d\l^i\prt_a\l_i-V^{cd}f_{de}V^{ef}f_{fa}
+{1\over2}V^{cd}f_{da}\,V^{ef}f_{fe})\nonumber\\
&&+\frac{1}{2\sqrt{2}}\phi\,V^{ab}\prt_a\l^i\prt_b\l_i+
V^{ab}(h_{ij}-b_{ij})\prt_b\l^i\prt_a\l^j
\nonumber\\
&&\left.+V^{ab}(h_{i(b}\prt_{a)}\l^i-b_{i[b}\prt_{a]}\l^i)V^{cd}f_{dc}
-
2V^{ab}(h_{i(b}\prt_{c)}\l^i-b_{i[b}\prt_{c]}\l^i)V^{cd}f_{da}frac{}{}\right)\,\, .
\labell{int2}
\eeqa
Finally, to compare  the couplings of three open string states and  massless
poles of  string amplitudes  with the corresponding
terms in the DBI theory, we will need also the  following action:
\beqa
\cL_{2,0}&=&-c\left(
{1\over2}V^{ab}\prt_a\l^i\prt_b\l_i-{1\over4} V^{ab}
f_{bc}V^{cd}f_{da}
+{1\over8}(V^{ab}f_{ba})^2\right)\nonumber\\
&=&-c\left(\frac{1}{2}(V_S)^{ab}\prt_a\l_i\prt_b\l^i-
\frac{1}{4}(V_S)^{ab}f_{bc}(V_S)^{cd}f_{da}
\right)\labell{int3}
\eeqa
where we have drop some total derivative terms in the second line above.

\section{From commutative to non-commutative variables}

Taking into account that the open string vertex operators correspond to non-commutative
gauge fields\cite{sw}, one should write the ordinary  open string fields  
 in terms of their non-commutative counterparts. 
In \cite{sw} the differential equation for non-commutative gauge field
was found  to be
\beqa
\delta\hF_{ab}(\theta)&=&\frac{1}{4}\delta\theta^{cd}\left(
2\hF_{ac}*\hF_{bd}+2\hF_{bd}*\hF_{ac}\right.
\nonumber\\
&&\left.-\hA_c*(\hD_d\hF_{ab}+\prt_d\hF_{ab})
-(\hD_d\hF_{ab}+\prt_d\hF_{ab})*\hA_c\right)+O(\hF^3)
\labell{delf}
\eeqa
where the field strength and $*$ product were defined to be
\beqa
\hF_{ab}&=&\prt_a\hA_b-\prt_b\hA_a-i\hA_a*\hA_b+i\hA_b*\hA_a\nonumber\\
&=&\prt_a\hA_b-\prt_b\hA_a-i[\hA_a,\hA_b]_M\nonumber\\
\hat{f}(x)*\hat{g}(x)&=&e^{\frac{i}{2}\theta_{ab}\prt^a_{x'}\prt^b_{x''}}
\hat{f}(x')\hat{g}(x'')|_{x'=x''=x}\,\, .
\labell{star}
\eeqa
Scattering amplitudes in the next section reproduce different couplings for finite
$\theta$. Therefore, to compare expected coupling of DBI action and string
amplitude we should integrate the differential equation \reef{delf} to find
relation between ordinary field strength, \ie $\hF_{ab}(\theta=0)$, and its 
non-commutative
counterpart, \ie $\hF_{ab}(\theta)$. We  take the integral in the  special path that
 $\theta_{ab}$
is proportional to a scalar, \ie $\theta_{ab}=\alpha\theta_{ab}$, and take
integral over $\alpha$ from $\alpha=0$ to $\alpha=1$. The result is
\beqa
F_{ab}&=&\hF_{ab}-\frac{1}{2}\theta^{cd}\left(
\hF_{ac}*''\hF_{bd}+\hF_{bd}*''\hF_{ac}
-\hA_c*''\prt_d\hF_{ab}
-\prt_d\hF_{ab}*''\hA_c\right)+O(\hA^3)
\nonumber
\eeqa
where now the non-commutative $*''$ product is defined to be
\beqa
\hat{f}(x)*''\hat{g}(x)&=&\frac{e^{\frac{i}{2}\theta_{ab}\prt^a_{x'}\prt^b_{x''}}-1}
{\frac{i}{2}\theta_{ab}\prt^a_{x'}\prt^b_{x''}}\hat{f}(x')\hat{g}(x'')|_{x'=x''=x}
\nonumber
\eeqa
to check the result, one may  differentiate  it to get equation \reef{delf} up
to order $O(\hA^3)$.
For abelian case that we are interested in, the transformation becomes
\beqa
F_{ab}&=&\hF_{ab}-\theta^{cd}\left(\hF_{ac}*'\hF_{bd}-\hA_c*'\prt_d\hF_{ab}\right)
+O(\hA^3)
\labell{fhf}
\eeqa
where the commutative $*'$ operates as
\beqa
\hat{f}(x)*'\hat{g}(x)&=&\frac{\sin(\frac{1}{2}\theta_{ab}\prt^a_{x'}\prt^b_{x''})}
{\frac{1}{2}\theta_{ab}\prt^a_{x'}\prt^b_{x''}}\hat{f}(x')\hat{g}(x'')|_{x'=x''=x}
\labell{star'}
\eeqa
Now if one compare equation \reef{fhf} for infinitesimal $\delta\theta$ and 
finite $\theta$, one may conclude  that to go from ordinary
product of two open string fields at $\theta=0$ to finite $\theta$ 
one should use the following
transformation as well:
\beqa
fg|_{\theta=0}&\longrightarrow &
f*'g|_{\theta\ne 0}
\labell{fg}
\eeqa
where $f$ and $g$ are any arbitrary open string fields. 
We will see in the next section that this multiplication rule is consistent 
with string theory scattering amplitudes of two open  and one closed string
states.

Now with the help of equation \reef{fhf} and \reef{fg}, one can write the 
DBI coupling \reef{int1}, \reef{int2} and \reef{int3} at $\theta=0$
 in terms of non-commutative fields at $\theta\ne0$
corresponding to open string vertex operator. In doing so, one should first
using \reef{fg} replace ordinary multiplication of two open string fields by 
the $*'$ multiplication. Then, using \reef{fhf}, the ordinary fields are shifted
to their non-commutative counterparts. For example, transformation of equation
\reef{int2} up to three open string states
becomes
\beqa
{\hat{\cL}_{2,1}}\!\!\!&=&\!\!\!-\ka c\left(({1\over2}\tau+V^{ab}(h_{ba}-b_{ba})+
\frac{1}{2\sqrt{2}}(\Tr(V)-4)\phi
)\right.\nonumber\\
&&\times({1\over2}V^{ab}\prt_a\hl^i*'\prt_b\hl_i-{1\over4} V^{ab}
\hf_{bc}*'V^{cd}\hf_{da}
+{1\over8}V^{ab}\hf_{ba}*'V^{cd}\hf_{dc})\nonumber\\
&&-V^{ab}(h_{bc}-b_{bc}+\frac{\phi\eta_{bc}}{2\sqrt{2}})
(V^{cd}\prt_d\hl^i*'\prt_a\hl_i-V^{cd}\hf_{de}*'V^{ef}\hf_{fa}
+{1\over2}V^{cd}\hf_{da}*'V^{ef}\hf_{fe})\nonumber\\
&&+\frac{1}{2\sqrt{2}}\phi\,V^{ab}\prt_a\hl^i*'\prt_b\hl_i+
V^{ab}(h_{ij}-b_{ij})\prt_b\hl^i*'\prt_a\hl^j
\nonumber\\
&&\left.+V^{ab}(h_{i(b}\prt_{a)}\hl^i-b_{i[b}\prt_{a]}\hl^i)*'V^{cd}\hf_{dc}
-
2V^{ab}(h_{i(b}\prt_{c)}\hl^i-b_{i[b}\prt_{c]}\hl^i)*'V^{cd}\hf_{da}\frac{}{}\right)\,\, .
\labell{int22}
\eeqa
We now  turn to string theory side and evaluate  these 
couplings using the conformal
field theory technique.
\section{Scattering Calculations}

In this section, we calculate various string scattering amplitudes. The 
amplitude describing scattering
of two closed strings  from D-brane with a background magnetic flux 
was calculated in \cite{mrg}.
There by analyzing the $t$-channel of the amplitude, we were able 
to find the linear
coupling of closed string fields to the D-brane and to show that they 
are consistent with
the coupling in
\reef{int0}.  
\subsection{Closed-Open couplings} \label{one}

Here, we wish to compare the  coupling of one massless 
non-commutative open string
field and one closed string field on the D-brane
to the results of the appropriate string couplings. In field theory, this 
coupling can be read from eq.~\reef{int1} and  the transformation \reef{fhf}.
In string theory side, on the other hand, this coupling is given by
the string scattering amplitude of 
one  open  and one closed string  state from the D-brane, that is,
\beqa
A^{\rm NS,NS-NS}&\sim&\int\,dx_1\,d^2z_2<V^{\rm NS}(k_1,\z_1,x_1)
\,V^{\rm NS-NS}(p_2,\veps_2,z_2,\bar{z}_2)>\,\, .
\labell{ans,nsns}
\eeqa
The details of the vertex operators appear in the Appendix. We already assumed in Sec. 2
that the closed string fields in the DBI action \reef{biact}
are independent of transverse coordinates. 
In  string theory side, it means that
the momentum of closed string vertex operators  have component only 
in the world-volume
directions, \ie $p_i=0$.
The techniques in calculating
the above string scattering amplitude may be found in 
refs.~\cite{aki,ours,scatd}.
The final result is
\beqa
A^{\rm NS,NS-NS}&=&\frac{\sqrt{T_p}\kappa c}{2}
\left(2k_{1a}(\cG\cdot\veps_2\cdot D\cdot\cG^T)^{\mu
a}\z_{1\mu}-2k_{1a}(\cG\cdot\veps_2\cdot D\cdot\cG^T)^{a\mu}\z_{1\mu}
\right.\nonumber\\
&&\qquad\qquad\qquad\qquad\qquad\qquad\qquad\qquad \left.-p_{2\mu}
(D\cdot\cG^T)^{\mu\nu}
\z_{1\nu}Tr(\veps_2\cdot D)\right)
\nonumber\\
A^{\rm NS,\tau}&=&\frac{\sqrt{T_p}\kappa c}{2}\z_1\cdot\cG\cdot p_2
\nonumber
\eeqa
where $D^\mu{}_\nu$ and $\cG^{\mu}{}_{\nu}$ matrices coming from closed 
and open
string vertices, respectively(see Appendix). We have also normalized 
the amplitudes at this
point by $-i\sqrt{T_p}\kappa c/2$, where $1/\sqrt{T_p}$, $\ka$ and $T_pc$ are
open string, closed string and D-brane coupling constants, respectively.
Substituting the appropriate polarizations for the open and closed string
 fields from the Appendix, 
one finds 
\beqa
A(\l,h)&=&\sqrt{T_p}\kappa c\left(\z_1\inn N\inn\veps_2\inn V^T\inn k_1
+k_1\inn V^T\inn\veps_2\inn N\inn\z_1\right)\nonumber\\
A(a,h)&=&-\sqrt{T_p}\kappa c\left(\z_1\inn V\inn\veps_2\inn V\inn k_1-
k_1\inn V\inn\veps_2\inn V\inn\z_1+k_1\inn V_A\inn\z_1\Tr(\veps_2\inn V^T)
\right)
\labell{aah}\nonumber\\
A(a,\phi)&=&-\frac{\sqrt{T_p}\kappa c}{2\sqrt{2}}\left(\z_1\inn V\inn V\inn k_1
-k_1\inn V\inn V\inn\z_1+k_1\inn V_A\inn\z_1
(\Tr(V)-4)\right)
\labell{aap}\nonumber\\
A(a,\tau)&=&-\frac{\sqrt{T_p}\kappa c}{2}k_1\inn V_A\inn\z_1
\labell{aat}\nonumber
\eeqa
where here and in the scattering amplitudes in subsequent sections $h$
stands for both graviton and Kalb-Ramond fields. 
In writing above equations, we have used the on-shell 
condition $k_1\inn V_S\inn\z_1=0$(see Appendix) and momentum conservation 
$k_1^a+p_2^a=0$. These terms are reproduced by the following action:
\beqa
{\hat{\cL}_{1,1}}&=&-\sqrt{T_p}\ka c\left(\frac{1}{2}V^{ab}\hf_{ba}
(\frac{1}{2}\tau+V^{ab}(h_{ba}-b_{ba})+\frac{1}{2\sqrt{2}}
(\Tr(V)-4)\phi)\right.\nonumber\\
&&\left.-V^{ab}(h_{bc}-b_{bc}+\frac{1}{2\sqrt{2}}\phi\eta_{bc})V^{cd}\hf_{da}
+2V^{ab}(h_{i(b}\prt_{a)}\hl^i
-b_{i[b}\prt_{a]}\hl^i)\right)
\,\, .
\labell{int11}
\eeqa
This is consistent with the DBI interaction \reef{int1} and the transformation
\reef{fhf} up to two open string fields. 
Note that transformation of $\prt_a\l$ can be read from dimensional reduction
of \reef{fhf}. 
\subsection{Open-Open-Open couplings}
Next,  we turn to the  coupling of three open string states. 
In string theory side this coupling is given by
\beqa
A^{NS,NS,NS}&\sim&\int dx_1dx_2dx_3 <V^{NS}(\z_1,k_1,x_1)\,
V^{NS}(\z_2,k_2,x_2)\,V^{NS}(\z_3,k_3,x_3)>
\nonumber
\eeqa
where the appropriate vertex operators are given in the Appendix. 
Using the world-sheet
conformal field theory, it is not difficult to perform the correlators 
above and
show that the integrand is invariant under $SL(2,R)$. Fixing 
this symmetry, one
finds 
\beqa
A&=&\frac{c\sin(\pi l)}{\pi\sqrt{T_p}}(k_1\inn\cG\inn\cG^T\inn\z_3\,
\z_1\inn\cG\inn\cG^T\inn\z_2+
k_3\inn\cG\inn\cG^T\inn\z_2\,\z_1\inn\cG\inn\cG^T\inn\z_3-
k_3\inn\cG\inn\cG^T\z_1\,\z_2\inn\cG\inn\cG^T\inn\z_3)
\nonumber
\eeqa
where we have defined $l\equiv -2k_1\inn V^T\inn \cF\inn V\inn k_2
=2k_1\inn V_A\inn k_2$ and
$V_A$ is antisymmetric part of the $V$ matrix \reef{vmatrix}.
We have also normalized the amplitude by the appropriate coupling 
factor $-c/2\pi\sqrt{T_p}$. 
The $\sin(\pi l)$ factor above arises basically from two different phase factors
corresponding to 
two distinct cyclic orderings of the vertex operators. Each phase factor steam
from the second terms of the world-sheet propagator \reef{pro2}.
Using polarization for
scalar and gauge field, one finds the following non-vanishing terms: 
\beqa
A(\l,\l,a)&=&\frac{c\sin(\pi l)}{\pi\sqrt{T_p}}\z_1\inn N\inn\z_2\,
k_1\inn V_S\inn\z_3
\nonumber\\
A(a,a,a)&=&\frac{c\sin(\pi l)}{\pi\sqrt{T_p}}(\z_1\inn V_S\inn\z_2\,
\z_3\inn V_S\inn
k_1+\z_2\inn V_S\inn\z_3\,\z_1\inn V_S\inn k_2+
\z_1\inn V_S\inn\z_3\,\z_2\inn V_S\inn k_3)
\labell{vect}
\eeqa
where $V_S$ is the symmetric part of the $V$ matrix \reef{vmatrix}. These
couplings may be  reproduced by 
\beqa
\hat{\cL}_{3,0}&=&\frac{ic}{8\pi\sqrt{T_p}}\left(
2V_S^{ab}\prt_a\hl_i*'[\ha_b,\hl^i]_M
-V_S^{ab}\hf_{bc}*'V_S^{cd}[\ha_d,\ha_a]_M\right)
\labell{int33}
\eeqa
where the Moyal bracket is defined in \reef{star} and the $*'$ operates on the
whole Moyal bracket. This fix the relation
between the
non-commutative two-form  in \reef{star} and the background metric and flux to be
\beqa
\theta^{ab}&=&4\pi V^{ab}_A\,\, .
\labell{theta}
\eeqa
The  DBI interaction \reef{int2}
and the transformations \reef{fg} and \reef{fhf}, reproduce various terms having two, three and
more open string fields. Terms which have three fields contain two different
parts. One part is just above action and the other part which have
three momentum may be verified to be zero. 
In action \reef{int33}, one may use , instead of $*'$, another 
multiplication rule, \eg ordinary multiplication \cite{sheikh} or $*$ 
multiplication \cite{sw}, all produce the same momentum space 
couplings\reef{vect}.
Hence,  although above
calculations of
three open string couplings can fix the non-commutative multiplication rule
in the definition of field strength,\ie $*$, 
it can not however uniquely fix the multiplication rule
between two open string field strengths. 

If one multiplies action \reef{int33} with a closed string field, say tachyon,
then the ordinary, $*$ and $*'$  and any other multiplication rule produce different 
momentum space couplings. In this case, string theory 
calculations can be used to 
fix uniquely the multiplication rule. 
To fix the multiplication rule between two open string fields, we
calculate string theory couplings of two open and one closed string states in the
momentum space.
These can be extracted from string scattering amplitudes of two open and one 
closed string states from the non-commutative D-brane  
which we now turn to calculate them.    

\subsection{Closed-Open-Open amplitudes} \label{quadi}

Scattering amplitude of one closed and two open string states can be
related to the appropriate  amplitude of four open string states in type 
I theory\cite{ours,aki}.
However, type I theory does not have open string tachyon, so the scattering
amplitude describing the decay of two massless open string 
to one closed string
tachyon in type 0 is not related to the known amplitude in type 
I theory. So we
explicitly calculate the tachyon amplitude, while using the idea in 
\cite{ours,aki} we
find the massless closed string amplitude from the known amplitude of type I
theory.
\subsubsection{Tachyon amplitudes}

The amplitudes
describing interaction of one closed string tachyon and two massless open
strings is given by
\beqa
A^{NS,NS,\tau}&\sim&\int\,dx_1dx_2d^2z<V^{NS}(\z_1,k_1,x_1)
V^{NS}(\z_2,k_2,x_2)
V^{\tau}(p_3,z_3)>
\nonumber
\eeqa
where the closed and open string vertex operators are given in the Appendix.
Here again using appropriate world-sheet propagators from \cite{ours}, 
one can evaluate the 
correlations above and show that the integrand in $SL(2,R)$ invariant.
Gauging this symmetry by fixing  $z_3=i$ and $x_2=\infty$, one arrives at
\beqa
A&\sim&2^{-2s-2}\int dx_1\left( (2s+1)\z_1\inn\cG\inn\cG^T\inn\z_2-
\frac{2i\z_1\inn\cG\inn D^T\inn p_3\,\z_2\inn\cG\inn p_3
}{x_1-i}+\frac{2i\z_1\inn \cG\inn p_3\,
\z_2\inn\cG\inn D^T\inn p_3}{x_1+i}\right)\nonumber\\
&&\qquad\qquad\qquad\times (x_1-i)^{s-l}(x_1+i)^{s+l}\nonumber
\eeqa
where the integral is taken from $-\infty$ to $+\infty$, and 
$s=-(p_3\inn V^T)^2=-2k_1\inn V_S\inn k_2$.
This integral is doable and the result is
\beqa
A&=&-\frac{i\ka c}{2}\left(a_1(s+l)-a_2(s-l)
\right)\frac{\Ga(-2s)}{\Ga(1-s-l)\Ga(1-s+l)}
\labell{Ansnst}
\eeqa
where $a_1$ and $a_2$ are two kinematic factors depending only on the 
space time momentum and polarization vectors
\beqa
a_1&=&-\z_1\inn\cG\inn D^T\inn p_3\z_2\inn\cG\inn p_3\nonumber\\
a_2&=&(s+l)\z_1\inn\cG\inn\cG^T\inn\z_2+
\z_1\inn\cG\inn p_3\z_2\inn\cG\inn D^T\inn p_3\,\, .
\nonumber
\eeqa
We have also normalized the amplitude \reef{Ansnst} at 
this point by the coupling factor $-i\ka c/{2\pi}$. 
A check of our calculations is that the amplitude \reef{Ansnst} 
satisfies the
Ward identity associated with the gauge invariance of the open 
string states, \ie
the amplitude vanishes upon substituting $\z_{ia}\rightarrow k_{ia}$.
This amplitude has the  pole structure 
at  $m_{open}^2=n/{\alpha'}$\footnote{We explicitly restore $\alpha'$ here.
Otherwise our conventions set $\alpha'=2$}.

\subsubsection{NS-NS amplitudes}

Next, we evaluate the amplitude describing the decay of two massless 
open NS strings
into one massless closed NSNS state. Using the idea in \cite{ours,aki}, we
relate this amplitude to amplitude of four massless open NS strings. 
Hence, we
 begins with the closed string amplitude which is given by
\beq
A\sim\int\,dx_1\,dx_2\,d^2z_3<V^{\rm NS}(k_1,\z_1,x_1)
\,V^{\rm NS}(k_2,\z_2,x_2)
\,V^{\rm NSNS}(p_3,\veps_3,z_3,\bar{z}_3)>\,\, .
\labell{ampall}
\eeq
If one evaluated the above correlators, one would find that the 
integrand is $SL(2,R)$ invariant.
Similar to the tachyon amplitude, the  appropriate 
way to fix this gauge would be to fix operators at 
$\{\bz_3,x_1,z_3,x_2\}=\{-i,y,i,\infty\}$. 

To relate the calculation here to that of a four point amplitude
of open superstrings\cite{jhs}, we write the latter  in $SL(2,R)$ 
invariant form, that is  
\beqa
A'&\sim&\int\,dx_1dx_2dx_3dx_4\,x_{12}^{4k_1\cdot k_2}x_{13}^{4k_1\cdot k_3}
x_{14}^{4k_1\cdot k_4}x_{23}^{4k_2\cdot k_3}x_{24}^{4k_2\cdot k_4}
x_{34}^{4k_3\cdot k_4}\nonumber\\
&&\times(\frac{a_1'}{x_{12}x_{13}x_{24}x_{34}}-
\frac{a_2'}{x_{13}x_{14}x_{23}x_{24}})
\labell{a1234}
\eeqa
where the kinematic factors $a_1'$ and $a_2'$ are 
\beqa
a'_1&=&4\{\z_1\inn\z_2\z_3\inn\z_4k_2\inn k_3+\z_1\inn\z_4 k_1\inn\z_2
k_4\inn\z_3+\z_2\inn\z_3k_2\inn\z_1k_3\inn\z_4
\nonumber\\
&&+\z_3\inn\z_4k_3\inn\z_2
k_4\inn\z_1
+\z_1\inn\z_2k_2\inn\z_3k_1\z_4-\z_1\inn\z_3k_1\inn\z_2k_3\inn\z_4
\nonumber\\
&&-\z_2\inn\z_4k_2\inn\z_1k_4\inn\z_3-\z_3\inn\z_4k_4\inn\z_2k_3\inn\z_1
-\z_1\inn\z_2k_1\inn\z_3k_2\inn\z_4 \}\ \ \ \ \ \ 
\labell{kinone}\\
a'_2&=&4\{k_2\inn k_4\z_1\inn\z_4\z_2\inn\z_3
+k_2\inn k_3\z_1\inn\z_3\z_2\inn\z_4-\z_1\inn\z_2\z_3\inn\z_4k_2\inn k_3
\nonumber\\
&&-\z_1\inn\z_4 k_1\inn\z_2k_4\inn\z_3-\z_2\inn\z_3k_2\inn\z_1k_3\inn\z_4
-\z_3\inn\z_4k_3\inn\z_2k_4\inn\z_1
\nonumber\\
&&-\z_1\inn\z_2k_2\inn\z_3k_1\z_4+\z_2\inn\z_3k_3\inn\z_1k_2\inn\z_4
+\z_1\inn\z_4k_4\inn\z_2k_1\inn\z_3
\nonumber\\
&&+\z_2\inn\z_4k_4\inn\z_1k_2\inn\z_3+\z_1\inn\z_3k_3\inn\z_2k_1\inn\z_4
\} \ \ .
\labell{kintwo}
\eeqa
where $\z_i$'s are the polarization of external states. Since we are interested
in this amplitude for transforming it to scattering amplitude of one closed and 
two open string states, we do not consider the phase factor associated with
the second term of  propagator
\reef{pro2}.
If one fixes  the $SL(2,R)$
symmetry by fixing the operators at $\{x_1,x_2,x_3,x_4\}=
\{-1,x,1,\infty\}$, one finds
\beqa
A'&\sim&(2)^{4k_1\cdot k_3-1}\int_{-1}^{+1}\,
(1+x)^{4k_1\cdot k_2}(1-x)^{4k_2\cdot k_3}
\nonumber\\
&& \times(\frac{a_1'}{1+x}-\frac{a_2'}{1-x})\,\, .
\labell{a1234two}
\eeqa
Now scattering amplitude \reef{ampall} can be read from 
amplitude \reef{a1234two}
by replacing 
\beqa
&&2k_1\rightarrow p_3\cdot D\,\,,\,\,k_2\rightarrow k_1\cdot \cG\,\,,\,\,
2k_3\rightarrow p_3\,\,,\,\,k_4\rightarrow k_2\cdot \cG\,\,,\,\,x\rightarrow iy
\nonumber\\
&&\z_{3\mu}\z_{1\nu}\rightarrow (\veps_3\inn D)_{\mu\nu}\,\,,\,\,\,
\z_{2\mu}\rightarrow \z_1\inn\cG_{\mu}
\,\,,\,\,\,\z_{4\mu}\rightarrow \z_2\inn\cG_{\mu}\,\,.\labell{trans}
\eeqa
Under these transformations
the amplitude \reef{a1234two} transforms to
\beqa
A&\sim&(2)^{-2s-1}\int_{-\infty}^{+\infty}\,(1+iy)^{s-l}(1-iy)^{s+l}
\nonumber\\
&&\times(\frac{a_1}{1+iy}-\frac{a_2}{1-iy})
\labell{ampalltwo}
\eeqa
and the kinematic factors \reef{kinone} and \reef{kintwo} become
\beqa
a_1&=&(s+l)\z_2\inn\cG\inn\veps_3\inn D\inn\cG^T\inn\z_1+
2k_2\inn\cG\inn\veps_3\inn D\inn\cG^T\inn\z_2\,
p_3\inn D\inn\cG^T\inn\z_1\nonumber\\
&&+2\z_1\inn\cG\inn\veps_3\inn D\inn\cG^T\inn k_1\,p_3\inn\cG^T\inn\z_2
+2\z_2\inn\cG\inn\veps_3\inn D\inn\cG^T\inn k_2 \,p_3\inn\cG^T\inn\z_1
\nonumber\\
&&+2k_1\inn\cG\inn\veps_3\inn D\inn\cG^T\inn\z_1\,p_3\inn D\inn\cG^T\inn\z_2-
\Tr(\veps_3\inn D)\,p_3\inn D\inn\cG^T\inn\z_1\,p_3\inn\cG^T\inn\z_2
\nonumber\\
&&-4k_2\inn\cG\inn\veps_3\inn D\inn\cG^T\inn k_1\,\z_2\inn\cG\inn\cG^T\z_1-
2\z_2\inn\cG\inn\veps_3\inn D\inn p_3\,k_2\inn\cG\inn\cG^T\inn\z_1
\nonumber\\
&&-2p_3\inn D\inn\veps_3\inn D\inn\cG^T\inn\z_1\,k_1\inn\cG\inn\cG^T\inn\z_2
\nonumber\\ 
a_2&=&-2s\,\z_1\inn\cG\inn\veps_3\inn D\inn\cG^T\inn\z_2+
(s+l)\Tr(\veps_3\inn D)\z_1\inn\cG\inn\cG^T\inn\z_2-
(s+l)\z_2\inn\cG\inn\veps_3\inn D\inn\cG^T\inn\z_1\nonumber\\
&&-2k_2\inn\cG\inn\veps_3\inn D\inn\cG^T\inn\z_2\,p_3\inn D\inn\cG^T\inn\z_1-
2\z_1\inn\cG\inn\veps_3\inn D\inn\cG^T\inn k_1\,p_3\inn\cG^T\inn\z_2
\nonumber\\
&&-2\z_2\inn\cG\inn\veps_3\inn D\inn\cG^T\inn k_2\,p_3\inn\cG^T\inn\z_1
-2k_1\inn\cG\inn\veps_3\inn D\inn\cG^T\inn\z_1\,p_3\inn D\inn\cG^T\inn\z_2
\nonumber\\
&&+2\z_1\inn\cG\inn\veps_3\inn  D\inn p_3\,k_1\inn\cG\inn\cG^T\inn\z_2+
2p_3\inn D\inn\veps_3\inn D\inn\cG^T\inn\z_2\,k_2\inn\cG\inn\cG^T\inn\z_1
\nonumber\\
&&+4k_1\inn\cG\inn\veps_3\inn D\inn\cG^T\inn k_2\,
\z_1\inn\cG\inn\cG^T\inn\z_2+
\Tr(\veps_3\inn D)p_3\inn\cG^T\inn\z_1\,p_3\inn D\inn\cG^T\inn\z_2\,\, .
\labell{a1a2}
\eeqa
The integral \reef{ampalltwo}
is doable and the result is the same as equation \reef{Ansnst} with above
kinematic factors. 
As a check of our calculations, we have inserted the 
dilaton polarization \reef{vdilaton} into the kinematic factor
and found that it is independent of the auxiliary vector $\ell^{\mu}$. 
Another check is that the amplitude satisfies the Ward identity associated with
the gauge invariance of the open string states.

\subsubsection{Massless poles}

Given the general form of the string amplitude in eq.~\reef{Ansnst}, one can
expand this amplitude as an infinite sum of 
terms reflecting the infinite tower
of open string states that propagate on the world-Volume of D-brane.
In the low energy domain, \ie $\alpha' m_{{\rm open}}<<1$, the first term
representing the exchange of massless string states dominates. In this case
the scattering amplitude \reef{Ansnst} reduces to
\beqa
A&=&\frac{i\ka c\sin(\pi l)}{4\pi s}(a_1+a_2)
+\cdots
\labell{Atau}
\eeqa
where dots represent contact terms and the infinite massive poles.
Making the appropriate explicit choices of polarizations, we find
\beqa
A_s(\l,\l,\tau)\!\!\!&=&\!\!\!\frac{i\ka c\sin(\pi l)}{8\pi s}
(l\z_1\inn N\inn \z_2)
+1\leftrightarrow 2
\nonumber
\\
A_s(a,a,\tau)\!\!\!&=&\!\!\!\frac{i\ka c\sin(\pi l)}{4\pi s}
(\frac{l}{2}\z_1\inn V_S\inn \z_2+
\z_1\inn V^T\inn p_3\,\z_2\inn V\inn p_3
)+1\leftrightarrow 2
\nonumber\\
A_s(\l,\l,\phi)\!\!\!&=&\!\!\!\frac{i\ka c\sin(\pi l)}
{8\pi\sqrt{2}s}(\frac{l}{2}(\Tr(D)+2)
-4k_1\inn V\inn V\inn k_2)\z_1\inn N\inn\z_2
+1\leftrightarrow 2
\labell{llt}\\
A_s(a, a,\phi)\!\!\!&=&\!\!\!\frac{i\ka c\sin(\pi l)}
{8\pi\sqrt{2}s}((\Tr(D)+2)
(p_3\inn V\inn\z_1 p_3\inn V^T\inn\z_2+\frac{l}{2}\z_1\inn V_S\inn\z_2)
\nonumber\\
&&+4k_1\inn V_S\inn\z_2(p_3\inn V\inn V\inn\z_1-\z_1\inn V\inn V\inn p_3)-
4k_1\inn V\inn V\inn k_2
\z_1\inn V_S\inn\z_2)+1\leftrightarrow 2
\nonumber\\
A_s(\l, a,h)\!\!\!&=&\!\!\!\frac{i\ka c\sin(\pi l)}{2\pi s}(
p_3\inn D\inn\veps_3\inn N\inn\z_1+
\z_1\inn N\inn\veps_3\inn D\inn p_3) k_1\inn V_S\inn\z_2
\nonumber\\
A_s(\l,\l, h)\!\!\!&=&\!\!\!\frac{i\ka c\sin(\pi l)}{4\pi s}(\frac{l}{2}
\Tr(\veps_3\inn D)
-4k_1\inn V\inn \veps_3^T\inn V\inn k_2)\z_1\inn N\inn\z_2+1\leftrightarrow 2
\nonumber\\
A_s(a,a,h)\!\!\!&=&\!\!\!\frac{i\ka c\sin(\pi l)}{4\pi s}
( \Tr(\veps_3\inn D)(
p_3\inn V\inn\z_1p_3\inn V^T\inn\z_2+\frac{l}{2}\z_1\inn V_S\inn\z_2)
-4k_1\inn V\inn\veps_3^T\inn V\inn k_2\z_1\inn V_S\inn\z_2\nonumber\\
&&\qquad\qquad\qquad\qquad+2(p_3\inn D\inn\veps_3\inn V^T\inn\z_2-
\z_2\inn V^T\inn\veps_3\inn D\inn p_3)
\z_1\inn V_S\inn k_2)+1\leftrightarrow 2 \,\, .
\nonumber
\eeqa
In writing explicitly the above massless poles,  
one finds some terms which is proportional to $s$ as well. We will
add these terms which have no contribution 
to the massless poles of field theory to 
the contact terms in
\reef{contact}. These amplitudes  should be reproduced in $s$-channel 
of field theory. We present
the calculation explicitly for the decay of two gauge fields into tachyon. 
This amplitude
can be evaluated in field theory as
\beqa
A'_s(a,a,\tau)&=&(\tV_{\tau a})^a(\tG_a)_{ab}(\tV_{aaa})^b
\labell{llt'}
\eeqa
where the propagator and the vertices can be read from 
\reef{int3}, \reef{int11} and \reef{int33}. They are 
\beqa
(\tG_a)^{ab}&=&\frac{i}{c}\frac{(V_S^{-1})^{ab}}{s}\nonumber\\
(\tV_{\tau a})^a&=&\frac{\sqrt{T_p}\ka c}{2}p_3\inn V_A^a\nonumber\\
(\tV_{aaa})^a&=&\frac{c\sin(\pi l)}{2\pi\sqrt{T_p}}\left(
\z_1\inn V_S\inn\z_2k_1-2k_1\inn V_S\inn\z_2\z_1\right)
\inn V_S^a+1\leftrightarrow 2\,\, .
\eeqa
In writing the above propagator from \reef{int3}, we have used the covariant
gauge $V_S^{ab}\prt_a \hA_b=0$. 
Replacing above propagator and vertices into \reef{llt'}, one finds exactly
the string massless pole $A_s(a,a,\tau)$. 
Similar calculations for the other open
string modes reproduces exactly the corresponding 
massless poles of the string
amplitudes \reef{llt}. 

Although the whole string scattering amplitude \reef{Ansnst} is
gauge invariant, vanishing upon substituting $\z_{ia}\rightarrow k_{ia}$, its massless
pole \reef{llt} is not gauge invariant which can be checked explicitly. 
So one expects that
the low energy contact terms of string amplitude  not to be gauge invariant either.
However, combination of the massless pole and contact terms  should be  gauge invariant. 
We turn now to evaluate these low energy contact terms of string amplitude \reef{Ansnst}

\subsubsection{Contact terms}

Having examined in detail the massless poles of string amplitudes, we now
extract the low energy contact terms of the string amplitude \reef{Ansnst}. 
Expanding the gamma
function appearing in this amplitude, one will find
\beqa
A&=&\frac{i\ka c}{2}\left((\frac{a_1+a_2}{2\pi})
\frac{\sin(\pi l)}{s}+(\frac{a_1-a_2}{2})
\frac{\sin(\pi l)}{\pi l}\right.\nonumber\\
&&\left.+(a_1+a_2)\frac{\sin(\pi l)}{\pi l}\,
\sum_{n=1}^{\infty}\zeta(2n+1)l^{(2n+1)}+k^2O(s,l)\right)
\,\, .
\labell{expand}
\eeqa
The factor $\sin(\pi l)/(\pi l)$ appears for all the contact terms. 
This indicates
that multiplication rule between  any two  open   strings 
is $*'$. However, as we will see in a moment, some of the terms in
the kinematic factors $a_1$ and $a_2$ are proportional to $l$. Hence, these terms have
overall factor of $\sin(\pi l)$ which produce the non-commutative multiplication rule
$*$ instead of $*'$ in field theory.
It is important to note that this higher order  derivative terms associated with
$*$ or $*'$ are
not the ones that
steams from massive pole of string amplitude. 
This factors  appear for
both the low  energy contact terms and the contact terms 
corresponding to massive
poles. 
Terms in the first line of \reef{expand} are the low energy 
massless pole and contact terms,
whereas the  
terms in the second line  are  effect of the massive 
poles of the string amplitude \reef{Ansnst}.

As anticipated above, not all the low energy contact terms are gauge invariant.
Therefore, we  separate the contact terms 
into gauge invariant
and gauge non-invariant terms. Moreover, we divide the gauge 
non-invariant terms into
two parts,  terms which have, apart from the overall factor, no momentum and two momenta. That is
\beqa
\frac{i\ka c\sin(\pi l)}{4\pi l}(a_1-a_2)&\equiv&A_c^M+A_c^{ng}+A_c^g
\labell{contact}
\eeqa
where $A_c^M$, $A_c^{ng}$ and $A_c^g$ are gauge non-invariant terms which 
have no momentum,  two momentum
and gauge invariant terms, respectively. The $A_c^M$ terms
are:
\beqa
A_c^M(a,a,\tau)&=&-\frac{i\ka c\sin(\pi l)}{4\pi}\z_1\inn V_A\inn\z_2
\nonumber\\
A_c^M(a,a,\phi)&=&\frac{i\ka c\sin(\pi l)}{4\pi\sqrt{2}}\left(
\z_1\inn V\inn V\inn\z_2-
\z_2\inn V\inn V\inn\z_1-(\Tr(D)+2)\z_1\inn V_A\inn\z_2\right)\nonumber\\
A_c^M(\l,a,h)&=&-\frac{i\ka c\sin(\pi l)}{2\pi}(
\z_1\inn N\inn\veps_3\inn V^T\inn\z_2
+\z_2\inn V^T\inn\veps_3\inn N\inn\z_1)\nonumber\\
A_c^M(a,a,h)&=&\frac{i\ka c\sin(\pi l)}{2\pi}(
\z_2\inn V^T\inn\veps_3\inn V^T\inn\z_1
-\z_1\inn V^T\inn\veps_3\inn V^T\inn\z_2-\frac{1}{2}
\Tr(\veps_3\inn D)\z_1\inn V_A\inn\z_2)\,\, .
\nonumber
\eeqa
In all above terms, the factor $\sin(\pi l)/(\pi l)$ reduces to $\sin(\pi l)$ 
which produces $*$ operator between two open string fields. In fact it is not difficult to see that above terms exactly 
reproduce by the following action:
\beqa
\hat{\cL}_{2,1}^M&=&\frac{i\ka c}{4\pi}\Tr\left(
\frac{1}{2}V^{ab}[\ha_b,\ha_a]_M
(\frac{1}{2}\tau+V^{ab}(h_{ba}-b_{ba})+\frac{1}{2\sqrt{2}}
(\Tr(V)-4)\phi)\right.\labell{int21}\\
&&\left.-V^{ab}(h_{bc}-b_{bc}+\frac{1}{2\sqrt{2}}\phi\eta_{bc})
V^{cd}[\ha_d,\ha_a]_M
+2V^{ab}(h_{i(b}[\ha_{a)},\hl^i]_M-b_{i[b}
[\ha_{a]},\hl^i]_M)\right)
\nonumber
\eeqa
where the non-commutative parameter of the Moyal bracket is the one appearing in \reef{theta}. 
Appearance of the Moyal bracket in \reef{int21} is 
consistent with the transformation \reef{fhf}.

The gauge non-invariant terms $A_c^{ng}$ are:
\beqa
A_c^{ng}(a,a,\tau)\!\!\!&=&\!\!\!\frac{i\ka c\sin(\pi l)}
{2\pi l}(\z_1\inn V_A\inn k_1\z_2\inn V_A\inn k_1)
+1\leftrightarrow 2\nonumber\\
A_c^{ng}(a,a,\phi)\!\!\!&=&\!\!\!\frac{i\ka c\sin(\pi l)}
{4\pi\sqrt{2}l}\left(
(\Tr(D)+2)k_1\inn V_A\inn\z_1 k_1\inn V_A\inn\z_2\right.\nonumber\\
&&\qquad\qquad\qquad\left. -2k_1\inn V_A\inn\z_2
(k_1\inn V\inn V\inn\z_1-\z_1\inn V\inn V\inn k_1)\right)+
1\leftrightarrow 2\labell{acng}\\
A_c^{ng}(\l,a,h)\!\!&=&\!\!\!\!\frac{i\ka c\sin(\pi l)}{2\pi l}\left(
2\z_1\inn N\inn\veps_3\inn V^T\inn k_1
+2k_1\inn V^T\inn\veps_3\inn N\inn\z_1\right)\z_2\inn V_A\inn k_1\nonumber\\
A_c^{ng}(a,a,h)\!\!\!&=&\!\!\!\frac{i\ka c\sin(\pi l)}{2\pi l}
\left(\Tr(\veps_3\inn D)k_1\inn V_A\inn
\z_1 k_1\inn V_A\inn\z_2\right.\nonumber\\
&&\qquad\qquad\qquad\left.-2k_1\inn V_A\inn\z_2
(k_1\inn V\inn\veps_3^T\inn V\inn\z_1-\z_1\inn V\inn\veps_3^T\inn V\inn k_1)
\right)+1\leftrightarrow 2 \,\, .
\nonumber
\eeqa
It is easy to check that, as expected,  under replacing $\z_{ia}\longrightarrow k_{ia}$
the non-zero terms of
 $A_c^M+A_c^{ng}$
cancels exactly the non-zero terms  of the massless poles in \reef{llt}. 
The factor $\sin(\pi l)/(\pi l)$ reproduces 
the $*'$ operator between two open string fields. Above momentum space couplings
are reproduced by the following action:
\beqa
\hat{\cL}_{2,1}^{ng}&=&-\ka c\Tr\left(\frac{}{}(V_A)^{ab}\ha_a*'\left(
\frac{1}{2}V^{cd}\prt_b \hf_{dc}(\frac{1}{2}\tau+V^{ef}(h_{fe}-b_{fe})+
\frac{1}{2\sqrt{2}}(\Tr(V)-4)\phi)\right.\right.
\nonumber\\
&&\left.\left.-V^{cd}(h_{de}-b_{de}+\frac{1}{2\sqrt{2}}
\phi\eta_{de})V^{ef}\prt_b \hf_{fc}
+2V^{cd}(h_{i(d}\prt_{c)}\prt_b\hl^i-b_{i[d}\prt_{c]}
\prt_b\hl^i)\right)\right)\,\, .\labell{lng}
\eeqa
Here also appearance of the non-commutative gauge field and derivative of
field strength is consistent with the transformation \reef{fhf}.

We turn now to the gauge invariant terms. The contact 
terms of two open string scalars
and one closed string are:
\beqa
A_c^g(\l,\l,\tau)&=&\frac{i\ka c\sin(\pi l)}{4\pi l}
(-s\z_1\inn N\inn\z_2)\nonumber\\
A_c^g(\l,\l,\phi)&=&\frac{i\ka c\sin(\pi l)}{8\pi\sqrt{2}l}\left(
\frac{}{}(\Tr(D)+2)(
-s\z_1\inn N\inn\z_2)
\right.\nonumber\\
&&\qquad\qquad\qquad\left.-4(s+k_1\inn V\inn V\inn k_2+
k_2\inn V\inn V\inn k_1)
\z_1\inn N\inn\z_2\right)\labell{allphi}\\
A_c^g(\l,\l,h)&=&\frac{i\ka c\sin(\pi l)}{4\pi l}\left(
\frac{1}{2}\Tr(\veps_3\inn D) 
(-s\z_1\inn N\inn\z_2)\right.\nonumber\\
&&\qquad\qquad\left.-4k_1\inn V\inn\veps_3^T\inn V\inn 
k_2\z_1\inn N\inn\z_2
-2(s-l)
\z_1\inn N\inn\veps_3\inn N\inn\z_2\right)
+1\leftrightarrow 2\,\, . \nonumber
\eeqa
These terms reproduce exactly by approperaite terms in \reef{int22}. This confirms the 
conjectured multiplication rule \reef{fg} between two open string fields.
Now the other 
gauge invariant contact terms are 
\beqa
A_c^g(\l,a,h)\!\!\!&=&\!\!\!\frac{i\ka c\sin(\pi l)}{2\pi l}\left(
2k_1\inn V_S\inn\z_2(
\z_1\inn N\inn\veps_3\inn V^T\inn k_2-
k_2\inn V^T\inn\veps_3\inn N\inn\z_1)\right.\nonumber\\
&&\left.+s(\z_1\inn N\inn\veps_3\inn V^T\inn\z_2-
\z_2\inn V^T\inn\veps_3\inn N\inn\z_1)
-2k_2\inn V_A\inn\z_2(k_1\inn V\inn\veps_3\inn N\inn\z_1+
\z_1\inn N\inn\veps_3\inn V\inn k_1)\right)\nonumber\\
A_c^g(a,a,\tau)\!\!\!&=&\!\!\!\frac{i\ka c\sin(\pi l)}{4\pi l}\left(
\z_1\inn V_A\inn k_1\z_2\inn V_A\inn k_2\right.\nonumber\\
&&\left.+\frac{l}{2}\z_1\inn V_A\inn\z_2-\frac{s}{2}\z_1\inn V_S\inn\z_2+
\z_1\inn V_A\inn k_2\z_2\inn V_A\inn k_1
-\z_1\inn V_S\inn k_2\z_2\inn V_S\inn k_1\right)\nonumber\\
A_c^g(a,a,\phi)\!\!\!&=&\!\!\!\frac{i\ka c\sin(\pi l)}{8\pi\sqrt{2}l}\left(
(\Tr(D)+2)\z_1\inn V_A\inn k_1\z_2\inn V_A\inn k_2-4k_1\inn V_A\inn
\z_1(k_2\inn V\inn V\inn\z_2-
\z_2\inn V\inn V\inn k_2)\right.\nonumber\\
&&+(\Tr(D)+2)(\frac{l}{2}\z_1\inn V_A\inn\z_2-\frac{s}{2}
\z_1\inn V_S\inn\z_2
+\z_1\inn V_A\inn k_2\z_2\inn V_A\inn k_1-\z_1\inn V_S\inn 
k_2\z_2\inn V_S\inn k_1)\nonumber\\
&&\left.+2s\z_1\inn V\inn V\inn\z_2-4\z_1\inn V_S\inn\z_2
k_1\inn V\inn V\inn k_2
+4k_2\inn V_S\inn\z_1(k_1\inn V\inn V\inn\z_2+\z_2\inn V\inn V\inn k_1)\right)
\labell{aag}\\
A_c^g(a,a,h)\!\!\!&=&\!\!\!\frac{i\ka c\sin(\pi l)}{4\pi l}\left(
\Tr(\veps_3\inn D)\z_1\inn V_A\inn k_1\z_2\inn V_A\inn k_2-
4k_1\inn V_A\inn\z_1(k_2\inn V\inn\veps_3^T\inn V\inn\z_2-
\z_2\inn V\inn\veps_3^T\inn V\inn k_2)\right.\nonumber\\
&&+\Tr(\veps_3\inn D)(\frac{l}{2}\z_1\inn V_A\inn\z_2-
\frac{s}{2}\z_1\inn V_S\inn\z_2
+\z_1\inn V_A\inn k_2\z_2\inn V_A\inn k_1-
\z_1\inn V_S\inn k_2\z_2\inn V_S\inn k_1)\nonumber\\
&&\left.+2s\z_1\inn V\inn \veps_3^T\inn V\inn\z_2-
4\z_1\inn V_S\inn\z_2k_1\inn V\inn\veps_3^T\inn V\inn k_2
+4k_2\inn V_S\inn\z_1(k_1\inn V\inn\veps_3^T\inn V\inn\z_2+
\z_2\inn V\inn \veps_3^T\inn V\inn k_1)\right)
\nonumber
\eeqa
plus $(1\leftrightarrow 2)$ for  equations that have two gauge fields. 
These gauge invariant terms are not fully consistent with the DBI terms
in \reef{int22}. However,
adding the following
terms to \reef{int22}, the resulting action  reproduces all the contact 
 terms in \reef{aag},
\beqa
\hat{\cL}_{2,1}^g&=&-\ka c\,\Tr\left(
{1\over2} V^{ab}\hf_{bc}*'(V_A)^{cd}\hf_{da}
({1\over2}\tau+V^{ab}(h_{ba}-b_{ba})+
\frac{1}{2\sqrt{2}}(\Tr(V)-4)\phi
)\right.\nonumber\\
&&-V^{ab}(h_{bc}-b_{bc}+\frac{1}{2\sqrt{2}}\phi\eta_{bc})
(V^{cd}\hf_{de}*'(V_A)^{ef}\hf_{fa}
)\nonumber\\
&&\left.+V^{ab}(h_{ib}-b_{ib})\prt_c\hl^i*'(V_A)^{cd}\hf_{da}
+(V_A)^{ab}(h_{ic}+b_{ic})\prt_b\hl^i*'V^{cd}\hf_{da}\right)\,\, .
\labell{int2new}
\eeqa
Now the string theory contact terms in  equations \reef{int11}, \reef{int21},
\reef{lng} and \reef{int2new} are  exactly  the DBI interactions
\reef{int1} in which  using the transformation \reef{fhf}  
with the normalization \reef{onormal}
its ordinary fields are written 
in terms of their non-commutative fields
up to three open string fields. This ends our illustration of
consistency between string theory scattering amplitudes and ordinary
DBI action in which
using transformations \reef{fhf} the ordinary open string fields are
written in terms of non-commutative fields.
\section{Discussion} \label{discuss}

Having integrated  the Seiberg-Witten differential equation
in a special path, we  write ordinary gauge fields in terms
of non-commutative fields for finite non-commutative two-form
parameter. We then use this change of variables to express the ordinary DBI
action in terms of non-commutative fields. We have also proposed a transformation
for multiplication of two arbitrary open string fields when one write
DBI action in terms of non-commutative variables. The resulting action was then 
compared with various world-sheet perturbative string
theory scattering amplitudes. We find completely agreement between the 
field theory
and string theory results. This indicates that the perturbative string theory
knows about the Seiberg-Witten differential equation.

Our calculations of string scattering amplitude of two massless 
open and one closed string
states confirmed our  proposed commutative multiplication rule \reef{fg}
between two  open string fields. 
It would be interesting
to perform the calculation of one closed and three open string states to find
transformation of multiplication rules between three open string states.

The scattering amplitudes considered in this paper fixed the relation between
ordinary and non-commutative fields up to three non-commutative fields. 
In principle, the perturbative string theory knows about all the terms in this
change of variable. So it would be interesting to extend 
our  method  to higher point functions 
to find other terms of the relation. 

In section 4.2 we reach to the conclusion that string scattering amplitude
\reef{vect} reproduce the action \reef{int33} in field theory. In \cite{sw}, 
Seiberg and Witten
conclude different action, \ie similar to  \reef{int33} with $*$ instead of
$*'$ operator.
These two  actions are identical up to some total derivative terms. In fact, using
the antisymmetric property of the non-commutative parameter, one finds
\beqa
 V_S^{ab}\hF_{bc}*'V_S^{cd}\hF_{da}&=& V_S^{ab}\hF_{bc}*V_S^{cd}\hF_{da}\nonumber\\
&=& V_S^{ab}\hF_{bc}V_S^{cd}\hF_{da}\nonumber
\eeqa
up to some total derivative terms.

Our calculations of string scattering amplitude of two open and one closed
string states from non-commutative D-brane are  also original. 
Using the two-dimensional conformal field theory, we performed calculations
for scattering amplitude of two  massless open string states and one closed
string tachyon explicitly. Whereas, using the idea that scattering amplitude of
open and closed string states can be read from approperaite amplitude of only
open string states\cite{ours,aki}, we were able to find an expression for scattering
amplitude of two open and one massless closed string states from non-commutative
D-branes.

\vspace{1cm}
{\bf Acknowledgments}

I would like to acknowledge useful conversation with R.C. Myers. I would also
like to thank ICTP for hospitality.
This work was supported by University of Birjand and IPM.

When I was finalizing this paper, the paper \cite{9909059} came out
which has some overlap with the results in section 4.

\newpage
\appendix
\section{ Perturbative string theory with background field}

In perturbative superstring theories, to study scattering amplitude of 
some external string states 
  in conformal
field theory frame,
one usually evaluate correlation function of their corresponding vertex
operators with
use of some  standard conformal field theory 
propagators \cite{pkllsw}.
In trivial flat background one uses an  appropriate 
{\it linear} $\sigma$-model to
derive the propagators and define the vertex operators.
 In
nontrivial
D-brane background the vertex operator remain unchanged while the standard
propagators need some modification. Alternatively, one may use a doubling
trick to convert the propagators to standard form and give the modification
to the vertex operators\cite{ours}. 
In this appendix we would like to consider a D-brane with constant gauge field 
strength / or antisymmetric Kalb-Ramond field in  all directions of the D-brane.
The modifications arising from the appropriate {\it linear} $\sigma$-model
 appear in the following boundary conditions \cite{leigh}\footnote{
Our notation and conventions follow  those established in \cite{ours}.
So we are working on the upper-half plane 
with boundary at $y=0$ which means $\prt_y$ is
normal derivative and $\prt_x$ is tangent derivative.
And our index conventions are that lowercase Greek
indices take values in the entire ten-dimensional
space-time, \eg $\mu,\nu=0,1,\ldots,9$; early Latin indices take values
in the world-volume, \eg $a,b,c=0,1,\ldots,p$; and middle Latin indices
take values in the transverse space, \eg $i,j =p+1,\ldots,8,9$.
Finally, our conventions
set $\ls^2=\alpha'=2$.}: 
\beqa
\prt_y X^a-i\cF^a{}_b\prt_x X^b\,\,=\,\,0&{\rm for }&a,b\,=0,1,\cdots p 
\nonumber\\
X^i\,\,=\,\,0&{\rm for}&i\,=p+1,\cdots 9
\labell{mixboundary}
\eeqa
where $\cF_{ab}$  are the  constant background fields, 
and these equations are imposed at $y=0$. 
The world-volume (orthogonal subspace) indices
are raised and lowered by $\eta^{ab}(N^{ij})$ 
and $\eta_{ab}(N_{ij})$, respectively.
Now we have to understand the modification of the conformal field theory
propagators arising from these mixed boundary conditions. To this end consider
the following general expression for propagator of $X^{\mu}(z,\bz)$ fields:
\beqa
<X^{\mu}(z,\bz)\,X^{\nu}(w,\bw)>
&=&-\eta^{\mu\nu}\log(z-w)-\eta^{\mu\nu}\log(\bz-\bw) \nonumber\\
&&-D^{\mu\nu}\log(z-\bw)-D^{\nu\mu}(\bz-w)
\labell{pro1}
\eeqa
where $D^{\mu\nu}$ is a constant matrix. 
To find this matrix, we impose
the boundary condition \reef{mixboundary} on the propagator \reef{pro1}, which 
yields
\beqa
\eta^{ab}-D^{ba}-\cF^{ab}-\cF^a{}_c D^{bc}&=&0\labell{D1}
\eeqa
for the world-volume directions, $D^{ij}=-N^{ij}$ for the orthogonal directions, and 
$D^{ia}=0$ otherwise.
Now equation \reef{D1} can be solved for $D^{ab}$, that is
\beqa
D_{ab}&=&2(\eta-\cF)^{(-1)}_{ab}-\eta_{ab}\labell{D2}\\
&=&2V_{ba}-\eta_{ab}
\eeqa
where matrix $V$ is the dual metric  that appears 
in the expansion of DBI action \reef{vmatrix}.
Note that the $D^{\mu\nu}$ is orthogonal matrix, 
\ie $D^{\mu}{}_{\alpha}D^{\nu\alpha}=\eta^{\mu\nu}$.  

Using two dimensional equation of motion, one can write 
the world-sheet fields in terms of right- and left-moving components. In terms of
these chiral fields,
closed NSNS and open NS vertex operators are
\beqa
V^{\rm NSNS}&=&:V_n(X(z),\psi(z),\phi(z),p):
:V_m(\tX(\bz),\tpsi(\bz),\tphi(\bz),p):
\nonumber\\
V^{\rm NS}&=&:V_n(X(x)+\tX(x),\psi(x)+\tpsi(x),\phi(x)+\tphi(x),k):\nonumber
\eeqa
where $\psi^{\mu}$ is super partner of world-sheet field $X^{\mu}$ and $\phi$
is world-sheet superghost field. The indices  $n,m$ refer to the superghost charge of vertex operators,
and $p$ and $k$ are closed and open string momentum, respectively.
In order to work with only right-moving fields, we use the following doubling
trick:
\beqa
\tX^{\mu}(\bz)\longrightarrow D^{\mu}{}_{\nu}X^{\nu}(\bz) &\,\,\,\,\,
\tpsi^{\mu}(\bz)\longrightarrow D^{\mu}{}_{\nu}\psi^{\nu}(\bz)&\,\,\,\,\,
\tphi(\bz)\longrightarrow \phi (\bz)\,\, .
\labell{trick}
\eeqa
These replacements in effect extend the right-moving 
fields to the entire complex plane and
shift modification arising from mixed boundary condition
from propagators to vertex operators. Under these replacement, world-sheet
propagator between all right-moving fields take the 
standard form \cite{mrg} except
the following boundary propagator:
\beqa
<X^{\mu}(x_1)\,X^{\nu}(x_2)>&=&-\eta^{\mu\nu}\log(x_1-x_2)+
\frac{i\pi}{2}\cF^{\mu\nu}\Theta(x_1-x_2)
\labell{pro2}
\eeqa
where $\Theta(x_1-x_2)=1(-1)$ if $x_1>x_2(x_1<x_2)$. Note that the orthogonal
property
of the $D$ matrix is an important ingredient for writing the propagators in the standard form. The vertex operators
under transformation \reef{trick} becomes
\beqa
V^{\rm NSNS}&=&V_n(X(z),\psi(z),\phi(z),p):
:V_m(D\inn X(\bz),D\inn \psi(\bz),\phi(\bz),p):
\nonumber\\
V^{\rm NS}&=&:V_n(X(x)+D\inn X(x),\psi(x)+D\inn\psi(x),2\phi(x),k):\,\, .
\nonumber
\eeqa
The vertex operator for closed string tachyon, massless NSNS and massless NS
states are
\beqa
V^{\rm \tau}&=&:V_n(p,z)::V_m(p\inn D,\bz):\nonumber\\
V^{\rm NSNS}&=&(\veps\inn D)_{\mu\nu}:V_n^{\mu}(p,z):
:V_m^{\nu}(p\inn D,\bz):\nonumber\\
V^{\rm NS}&=&(\z\inn\cG)_{\mu}:V_n^{\mu}(2k\inn V^T,x):\nonumber
\eeqa
where $\cG^{ab}=(\eta^{ab}+D^{ab})/2=V^{ba}$ for gauge field, 
$\cG^{ij}=(\eta^{ij}-D^{ij})/2=N^{ij}$
for scalar field and $\cG^{ai}=0$ otherwise.
The open string  vertex operators in $(0)$ and $(-1)$ pictures are
\beqa
V^{\mu}_0(k,x)&=&\left(\prt X^{\mu}(x)+ik\inn\psi(x)\,
\psi^{\mu}(x)\right)e^{i k\cdot X(x)}\nonumber\\
V^{\mu}_{-1}(k,x)&=&e^{-\phi(x)}\psi^{\mu}(x)e^{i k\cdot X(x)}\nonumber\\
V_0(k,x)&=&ik\inn\psi(x)e^{ik\cdot X(x)}\nonumber\\
V_{-1}(k,x)&=&e^{-\phi(x)}e^{ik\cdot X(x)}\,\, .
\nonumber
\eeqa
The physical conditions for the massless open string are
\[
k\inn V_S\inn k=0\,\,\,\,\,,\,\,\,\,\, k\inn V_S\inn\z=0
\]
and for massless closed string are $p^2=0$ and $p_{\mu}\veps^{\mu\nu}=0$ where
$\veps$ is the closed string polarization which is 
traceless and symmetric(antisymmetric)
for graviton(Kalb-Ramond) and 
\beq
\veps^{\mu\nu}=\frac{1}{\sqrt{8}}(\eta^{\mu\nu}-\ell^{\mu}p^{\nu}
-\ell^{\nu}p^{\mu})\,\,\,\,\,,\,\,\,\,\, \ell\inn p=1
\labell{vdilaton}
\eeq
for the dilaton. Using the fact that $D^{\mu\nu}$ is orthogonal
matrix, one finds the following identities:
\[
\cG\inn\cG^T=\cG^S\,\,\,,\,\,\,(D\inn\cG^T)^{ab}=\cG^{ab}\,\,\,,\,\,\,
(D\inn\cG^T)^{ij}=-N^{ij}
\]
where the $\cG^S$ is symmetric part of the $\cG$ matrix.

\end{document}